\documentclass[prb, twocolumn]{revtex4-1}

\usepackage{amsmath}
\usepackage{amsfonts}
\usepackage{graphicx}
\usepackage{hyperref}

\begin{document}

\title{Quantum matrix diagonalization visualized}

\author{Kevin Randles}
\email{kevinrandles@mail.weber.edu}
\author{Daniel V. Schroeder}
\email{dschroeder@weber.edu}
\affiliation{Department of Physics, Weber State University, Ogden, UT 84408-2508}

\author{Bruce R. Thomas}
\email{bthomas@carleton.edu}
\affiliation{Department of Physics and Astronomy, Carleton College, Northfield, MN 55057}

\begin{abstract}

We show how to visualize the process of diagonalizing the Hamiltonian matrix to find the energy eigenvalues and eigenvectors of a generic one-dimensional quantum system.  Starting in the familiar sine-wave basis of an embedding infinite square well, we display the Hamiltonian matrix graphically with the basis functions alongside.  Each step in the diagonalization process consists of selecting a nonzero off-diagonal matrix element, then rotating the two corresponding basis vectors in their own subspace until this element is zero.  We provide Mathematica code to display the effects of these rotations on both the matrix and the basis functions.  As an electronic supplement we also provide a JavaScript web app to interactively carry out this process.

\end{abstract}

\maketitle

\section{Introduction}

A number of recent papers in educational physics journals have highlighted the usefulness of matrix methods for solving the time-independent Schr\"odinger equation.  Applications have included one-dimensional potential wells,\cite{Marsiglio} double wells,\cite{Jelic,Dauphinee} spherical potentials,\cite{Jugdutt} periodic potentials,\cite{Pavelich,LeVot,Pavelich2} the helium atom,\cite{Hutchinson,MasseWalker} and other systems of two or three interacting particles.\cite{Na}

Of course matrix methods date back to the birth of quantum mechanics.  The new development is the ever-increasing availability of personal computers equipped with easy-to-use software\cite{Mathematica,Maple,Matlab,SciPy} that can diagonalize large matrices almost instantly.  Anyone who understands basic linear algebra can now use this software to solve quantum systems that could previously be tackled only by dedicated researchers.

We typically treat these miraculous diagonalization routines as black boxes, but there are good pedagogical reasons to peek inside and see how they work---or at least how they \textit{could} work.  One reason is that students are naturally curious about how things work.  Most students of quantum mechanics have taken only a first course in linear algebra and have therefore learned only one matrix diagonalization algorithm:  solve the characteristic polynomial for the eigenvalues, then plug these back into the eigenvalue equation and solve the resulting linear systems for the eigenvectors, one by one.  But this process is practical only for very small matrices.  How, then, do computers diagonalize matrices with hundreds, or even thousands, of rows and columns?

Besides satisfying students' curiosity, looking inside a practical matrix diagonalization algorithm can help build students' geometrical intuition for the high-dimensional vector spaces in which quantum states live.  We express these states, and the operators that act on them, with respect to a particular basis.  When we diagonalize a Hamiltonian matrix we are rotating our basis vectors.  In a two-dimensional vector space this rotation is confined to a single plane, but in larger vector spaces the number of independent rotation planes grows roughly in proportion to the dimension squared.

In this paper we describe a diagonalization algorithm (originally due to Jacobi\cite{jacobi}) that consists of successive rotations of basis vectors.  We then show how to implement the algorithm in a visual way, allowing students to see the effects of each individual rotational step on both the Hamiltonian matrix and the basis functions.  We provide Mathematica code for this implementation in Fig.~\ref{fig:MathematicaCode}, and we provide a JavaScript web app version as an electronic supplement.\cite{supplement}  Our hope is that this software will make the connection between wavefunctions and matrix representations more vivid for students, and help them understand the diagonalization process not as a mere symbol-pushing calculation but as a geometrical transformation.

\section{Putting the problem into matrix form}

Consider a quantum particle in one dimension, trapped in a potential $V(x)$ of arbitrary shape such that its low-lying energy eigenstates are localized between $x=0$ and $x=a$.  To find these energy eigenstates $\psi_n(x)$ and their corresponding energies $E_n$ we must solve the time-independent Schr\"odinger equation,
\begin{equation}\label{TISE}
	\hat{H} \psi_n = E_n \psi_n,
\end{equation}
where $\hat{H}$ is the Hamiltonian operator,
\begin{equation}
\hat{H} = -\frac{\hbar^2}{2M}\frac{d^2}{dx^2} + V(x).
\end{equation}

We proceed by introducing a set of normalized basis functions,
\begin{equation}
\varphi_m(x) = \sqrt{\frac{2}{a}}\sin\Bigl(\frac{m\pi x}{a}\Bigr),
\qquad m = 1, 2, \ldots,
\end{equation}
conveniently chosen to be eigenstates of the kinetic energy term of the Hamiltonian---that is, eigenstates of a hypothetical infinite square well extending from 0 to~$a$.  We can then expand each low-lying eigenstate in terms of these basis functions:
\begin{equation}\label{psiExpansion}
	\psi_n = \sum_m c_{mn} \varphi_m,
\end{equation}
where the first subscript on the coefficient $c_{mn}$ indicates which component, while the second subscript indicates which eigenstate.  Inserting this expansion on both sides of Eq.~(\ref{TISE}) and then taking the inner product with an arbitrary basis function $\varphi_l$, we obtain
\begin{equation}\label{FouriersTrick1}
	\sum_m \langle \varphi_l | \hat{H} | \varphi_m \rangle c_{mn}
	=  E_n \sum_m c_{mn} \langle \varphi_l | \varphi_m \rangle = E_n c_{ln}.
\end{equation}
The inner product on the left defines the $lm$ \textit{matrix element} of $\hat H$,
\begin{equation}\label{HamIP}
H_{lm} = \langle \varphi_l | \hat{H} | \varphi_m \rangle = \int_0^a\!\varphi_l(x)\hat{H}\varphi_m(x)\,dx,
\end{equation}
so Eq.~(\ref{FouriersTrick1}) is simply
\begin{equation}
	\sum_{m=1}^\infty H_{lm} c_{mn} = E_n c_{ln},
\end{equation}
which in matrix notation reads
\begin{equation}\label{matrixformTISE}
	\begin{bmatrix}
		H_{11} & H_{12} & \cdots \\
		H_{21} & H_{22} & \cdots \\
		\vdots & \vdots & \ddots \\
	\end{bmatrix}
	\begin{bmatrix}
	c_{1n} \\
	c_{2n} \\
	\vdots
	\end{bmatrix}
	= E_n 
	\begin{bmatrix}
	c_{1n} \\
	c_{2n} \\
	\vdots
	\end{bmatrix}.
\end{equation}

Solving this matrix eigenvalue equation is equivalent to solving the original operator eigenvalue equation, Eq.~(\ref{TISE}).  Because the basis functions $\varphi_m(x)$ are real, the Hamiltonian matrix is real and symmetric.  In numerical solutions one must truncate the matrix to a finite number of rows and columns, but this truncation is rarely a practical hindrance to solving one-dimensional problems.  The most computationally intensive part of the calculation is not solving the eigensystem but rather computing the matrix elements in the first place, using Eq.~(\ref{HamIP}).\cite{remark1}  The method outlined above can also be readily generalized to other sets of basis functions and to multidimensional problems.\cite{Pavelich,Pavelich2,Hutchinson,MasseWalker,Na}

Without loss of generality we can choose the eigenvectors of Eq.~(\ref{matrixformTISE}) to be purely real and normalized.  Then these column vectors form an orthogonal matrix:
\begin{equation}\label{cMatrix}
C = \begin{bmatrix}
		c_{11} & c_{12} & \cdots \\
		c_{21} & c_{22} & \cdots \\
		\vdots & \vdots & \ddots \\
	\end{bmatrix}.
\end{equation}
Performing a similarity transformation with $C$ diagonalizes the Hamiltonian:
\begin{equation} \label{CDiagonalizesH}
C^T H \, C = \begin{bmatrix}
		E_1 & 0 & \cdots \\
		0 & E_2 & \cdots \\
		\vdots & \vdots & \ddots \\
	\end{bmatrix}.
\end{equation}

\section{Jacobi rotations}

We now turn to the problem of solving the eigensystem of Eq.~(\ref{matrixformTISE}).  Our method,\cite{NR} originally due to Jacobi,\cite{jacobi} consists of performing a succession of rotations in two-dimensional subspaces of the (truncated) high-dimensional vector space, with each rotation chosen to zero out a single off-diagonal element of~$H$.

\begin{figure*}[t!]
\includegraphics[width=6.6in]{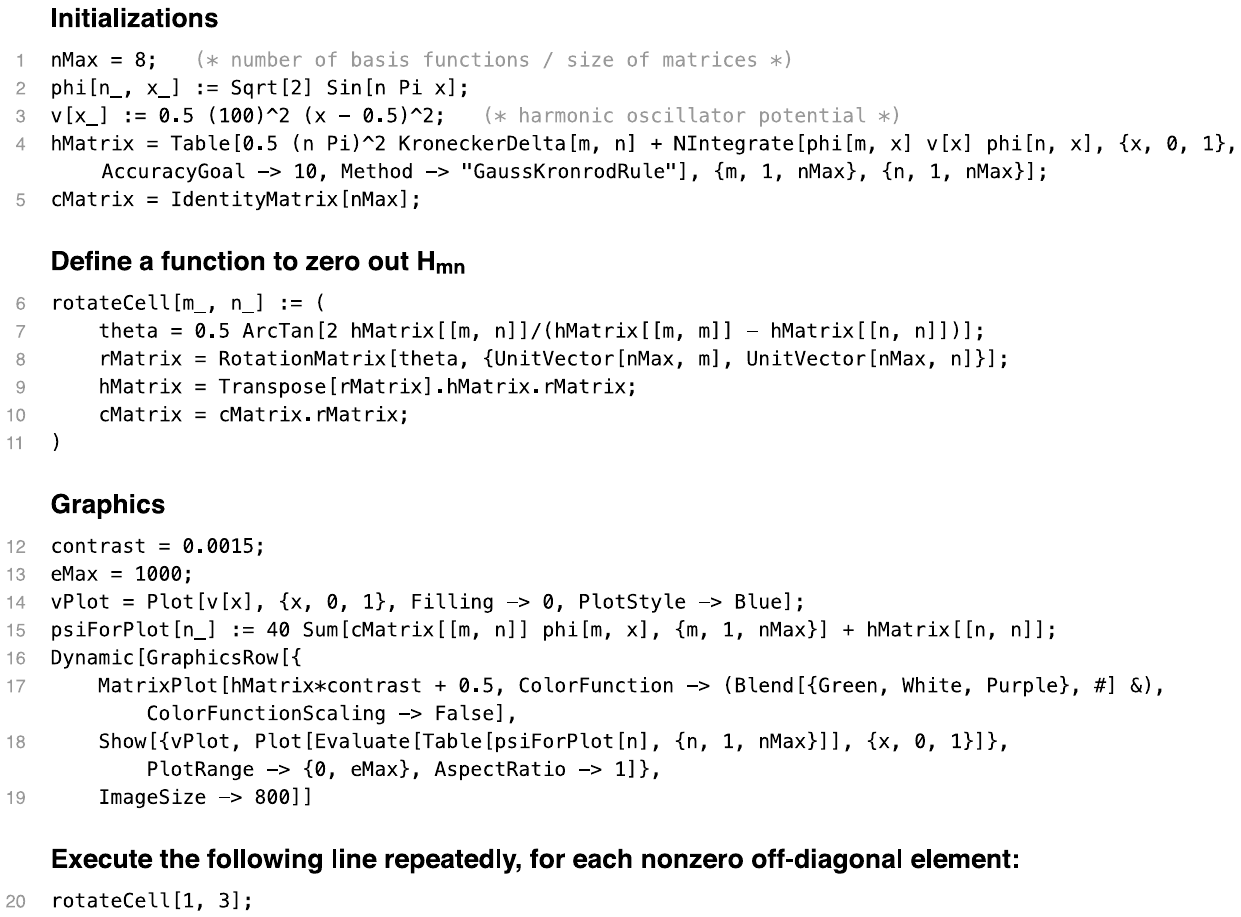}
\caption{Mathematica code to implement the Jacobi rotation algorithm, with dynamically updated displays of the Hamiltonian matrix and the corresponding basis functions.}
\label{fig:MathematicaCode}
\end{figure*}

We accomplish each rotation with a matrix of the form
\begin{equation}\label{rotationMatrix}
R = \begin{bmatrix}
\,1 \\[-5pt]
& \ddots \\
&& \cos\theta & \cdots & {-}\sin\theta \\[-3pt]
&& \vdots & 1 & \vdots \\
&& \sin\theta & \cdots & \ \cos\theta \\[-3pt]
&&&&& \ddots \\
&&&&&& 1\, \\
\end{bmatrix},
\end{equation}
consisting of an identity matrix except for the four entries that accomplish a rotation by $\theta$ in a particular plane; let us say these entries are in rows (and columns) $m$ and~$n$.  Under this rotation, the Hamiltonian transforms according to
\begin{equation}\label{HRotate}
H_\textrm{new} = R^T H_\textrm{old} R,
\end{equation}
and we will choose the angle $\theta$ so that the $mn$ element of $H_\textrm{new}$ is zero.  We then repeat this process with different choices of $m$ and $n$, zeroing out the off-diagonal elements of $H$ in succession.  Subsequent rotations will again make $H_{mn}$ nonzero, because each transformation changes all the elements in rows (and columns) $m$ and~$n$.  Still, the off-diagonal elements of $H$ get smaller and smaller as we repeatedly cycle through all of the possible $mn$ pairs.  (See Ref.~\onlinecite{NR} for a proof of convergence.)  We continue until $H$ is diagonal to our desired precision.

Once we are finished, the diagonal elements of $H$ will be the eigenvalues we seek.  Meanwhile, the product of all the rotation matrices will be (approximately) equal to the matrix $C$ of Eqs.\ (\ref{cMatrix}) and (\ref{CDiagonalizesH}):
\begin{equation}\label{CFromRs}
C = R_1 R_2 R_3 \cdots,
\end{equation}
where the subscripts indicate the sequence of the rotations.  We can then use Eq.~(\ref{psiExpansion}) to construct the eigenfunctions.  In fact, we can (and will) use successive approximations to $C$ to obtain approximate eigenfunctions, with ever-increasing accuracy, at each stage during the process.

To find the correct angle $\theta$ to zero out the new value of $H_{mn}$, it suffices to write a simplified version of the right-hand side of Eq.~(\ref{HRotate}) for just the rotation's two-dimensional subspace:
\begin{equation}
\begin{bmatrix} \cos\theta & \sin\theta \\ {-}\sin\theta & \cos\theta \\ \end{bmatrix}
\begin{bmatrix} H_{mm} & H_{mn} \\ H_{nm} & H_{nn} \\ \end{bmatrix}
\begin{bmatrix} \cos\theta & {-}\sin\theta \\ \sin\theta & \cos\theta \\ \end{bmatrix},
\end{equation}
where $H_{nm} = H_{mn}$ because $H$ is symmetric.  Working out either off-diagonal element of this product and setting the result to zero gives
\begin{equation}\label{rotationAngle}
	\theta = \frac{1}{2} \tan^{-1} \biggl( \frac{2H_{mn}}{H_{mm} - H_{nn}} \biggr).
\end{equation}

The order in which we choose $mn$ pairs for Jacobi rotations is not critical.  For ``manual'' diagonalization as we describe below, it is natural to begin with the larger off-diagonal elements of~$H$.  Alternatively, one can simply iterate through the off-diagonal elements in order by row and column.  Both the Mathematica code that we present below and the JavaScript web app that we provide as a supplement leave the order up to the user.

\section{Implementation in Mathematica}

Figure \ref{fig:MathematicaCode} shows an implementation of the Jacobi rotation algorithm in Mathematica.  The code uses natural units in which $\hbar = M = a = 1$.

The first line of code sets the number of basis functions used in the expansion (Eq.~(\ref{psiExpansion})), which is then the dimension of every vector and matrix.  The value 8 is large enough to get the main ideas across without requiring more than a few dozen rotational steps; for increased accuracy we have used values up to around~20.

The next two lines define the sinusoidal basis functions and the potential energy function.  For demonstration purposes we use a harmonic oscillator potential, centered at $x=0.5$, with a classical angular frequency of 100 in natural units (sufficient to confine the low-lying energy eigenfunctions within the interval from 0 to~1).

Line 4 generates the Hamiltonian matrix by breaking up the inner product of Eq.~(\ref{HamIP}) into kinetic and potential energy terms: 
\begin{equation}\label{HmnPieces}
	H_{mn} = \frac{1}{2} n^2 \pi^2 \delta_{nm} + \langle \varphi_m | \hat{V} | \varphi_n \rangle.
\end{equation} 
The first term is just the eigenvalues of the embedding infinite square well, whose eigenfunctions are $\varphi_n(x)$.  The \texttt{Method} and \texttt{AccuracyGoal} options speed up the numerical integration somewhat, and eliminate warning messages about matrix elements that are zero due to symmetry.

Line 5 initializes the matrix $C$ of Eq.~(\ref{cMatrix}) to the identity matrix, so we can later multiply it by each rotation matrix according to Eq.~(\ref{CFromRs}).

Lines 6--11 define a function that performs a single Jacobi rotation to zero out a selected off-diagonal cell~$H_{mn}$.  This function first finds the rotation angle using Eq.~(\ref{rotationAngle}), then sets up the rotation matrix in the form of Eq.~(\ref{rotationMatrix}).  Line 9 rotates the Hamiltonian matrix according to Eq.~(\ref{HRotate}), while line 10 updates the $C$ matrix according to Eq.~(\ref{CFromRs}).  For the matrix products we simply use Mathematica's \texttt{.}\ operator, making no attempt to optimize for the sparse rotation matrix.  (A much more efficient implementation is given in Ref.~\onlinecite{NR}.)

\begin{figure}[t!]
\includegraphics[width=8.6cm]{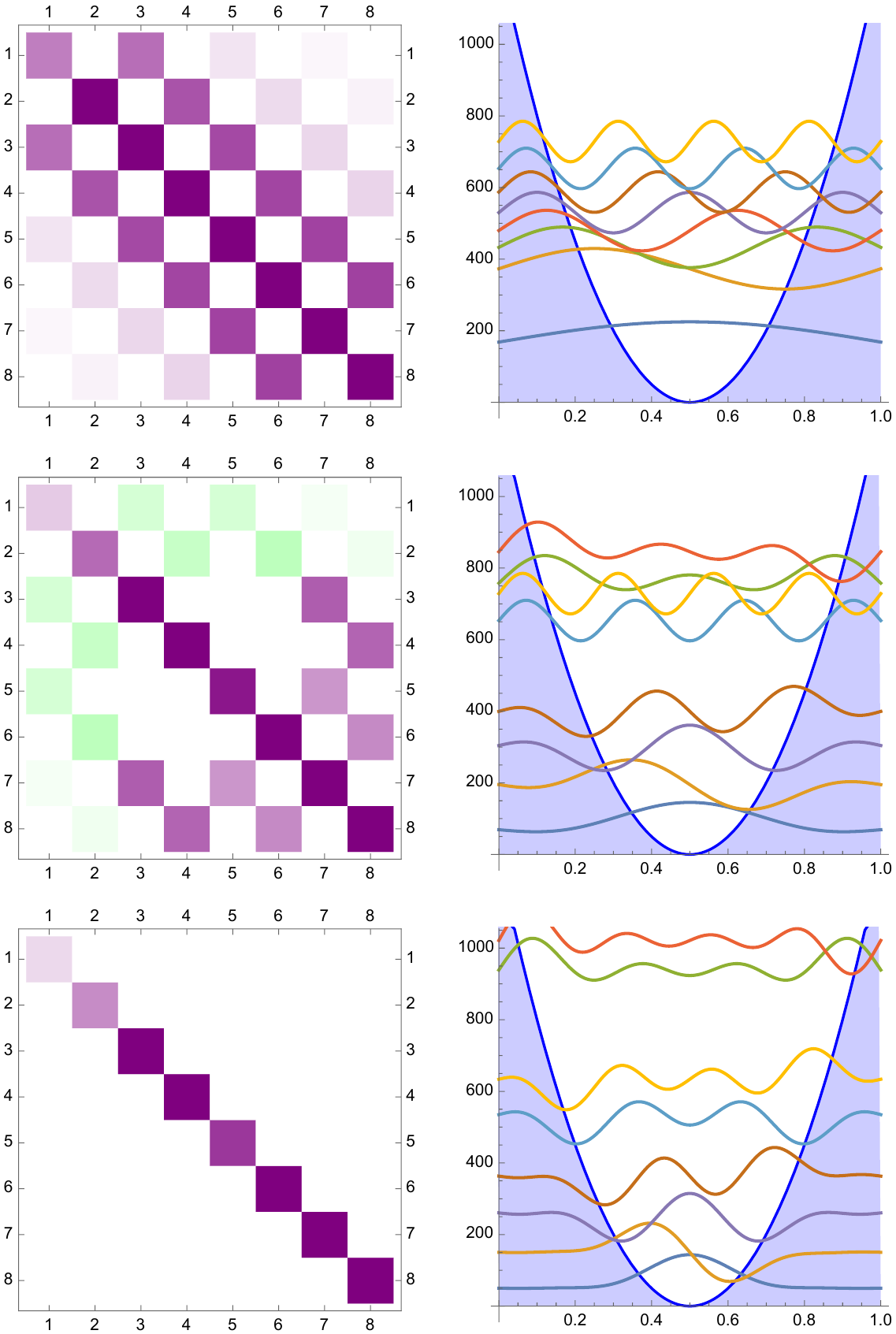}	
\caption{Mathematica graphical output illustrating the diagonalization process for a harmonic oscillator potential.  At each stage we show the Hamiltonian matrix at left (with darker shades for larger-magnitude elements and zero represented by white), and a plot of $V(x)$ and the eight basis functions at right.  The top images show the initial Hamiltonian and  basis functions, while the middle and bottom images show the results after performing 4 and 24 rotations, respectively.  Notice in the final image that the low-lying energies have approximately equal spacing, as expected for a quantum harmonic oscillator.}
\label{fig:MathematicaOutput}
\end{figure}

Lines 12--19 create our visualization scheme: a matrix plot of the Hamiltonian alongside a plot of the current basis functions overlaid on the potential (see Fig.~\ref{fig:MathematicaOutput}).  The \texttt{Dynamic} function causes these plots to update automatically each time \texttt{rotateCell} is executed.  Lines 12 and 13 define two variables that should be adjusted from time to time as needed:  a contrast for the matrix plot and a maximum energy for the potential/eigenfunction plot.  Line 15 scales the wavefunctions by an arbitrary factor and shifts each of them upward by its (average) energy.  Line 17 plots the Hamiltonian using a color scheme that maps zero to white, with distinct colors (modify according to taste!) for positive and negative values.

After executing lines 1 through 19, one merely has to repeatedly execute line 20, calling the \texttt{rotateCell} function for each Jacobi rotation.  The function's two arguments are the row and column numbers of the off-diagonal element (actually two symmetrically located elements) of $H$ that will be zeroed out.  The matrix display will update to show the newly zeroed elements, along with side effects on the other elements in the same rows and columns.  Meanwhile, the plots of the two corresponding basis functions will update to reflect their newly mixed shapes as well as their new average energies.  Figure~\ref{fig:MathematicaOutput} shows the graphical display at three different stages during the diagonalization procedure.

The harmonic oscillator potential is a useful first example because we can compare to the known exact solutions.  The algorithm can handle arbitrary potential shapes.  Some interesting examples are explored in the exercises in Sec.~\ref{exerciseSection}. 

After calling \texttt{rotateCell} two or three times for each nonzero off-diagonal element, it may be necessary to increase the \texttt{contrast} variable in order to distinguish the colors of the now-small off-diagonal elements from white.  Alternatively, it may be more convenient to simply display the matrix numerically, with an instruction such as
\begin{verbatim}
  Dynamic[MatrixForm[Chop[hMatrix]]]
\end{verbatim}
(where the \texttt{Chop} function sets very small values to zero to simplify the display).

As a test, one can check that the instruction
\begin{verbatim}
  Eigensystem[hMatrix]
\end{verbatim}
yields the same eigenvalues and eigenvectors as the Jacobi rotation process.  Going through so many steps to accomplish something Mathematica can do in a single line might seem foolish, but we feel that the visual step-by-step process provides insight that one cannot obtain from a black-box function call.

\section{Implementation as a web app}

As an electronic supplement\cite{supplement} we provide an alternate visual implementation of the Jacobi rotation algorithm in the form of a JavaScript web app.  This version has the drawback of making the code less accessible and harder to modify.  On the other hand, its graphical user interface is more intuitive and appealing.  Instead of typing the row and column numbers of a chosen off-diagonal matrix element, the user simply clicks or taps that element in the visual display.  The two corresponding basis functions are immediately highlighted.  Then the user presses and turns a graphical dial control by any desired angle, as both the matrix display and the basis functions update in real time to show the effects of the basis vector rotation.  To facilitate diagonalizing larger matrices, the interface also provides shortcut buttons to select the largest remaining off-diagonal element and to rotate by the correct angle to zero out the selected element.

The web app provides a menu of several built-in potential functions, with adjustable parameters.  There is also an option to draw an arbitrary potential function.

\section{Student exercises}
\label{exerciseSection}

We hope that students will be curious enough to use the code we have provided for open-ended exploration, asking and answering their own questions.  In case further direction might be helpful, here are some suggested exercises.

\begin{enumerate}

\item Derive Eq.~(\ref{rotationAngle}) for the Jacobi rotation angle.

\item Explain the checkerboard appearance of the harmonic oscillator Hamiltonian matrix in Fig.~\ref{fig:MathematicaOutput}.

\item Explain why the initial value of $H_{13}$ for the harmonic oscillator Hamiltonian matrix is positive.  Suppose that you choose $H_{13}$ as the first matrix element to zero out via a Jacobi rotation.  What is the rotation angle in degrees?  What is the rotation matrix?  Explain why the rotation causes the average energy of the $n=1$ basis state to decrease.

\item When you perform a Jacobi rotation on basis vectors $m$ and $n$, \textit{all} the matrix elements in the corresponding rows and columns can (potentially) change.  Why, then, don't all the other basis \textit{functions} change?  Explain carefully.

\item Suppose that you're in a hurry to find the ground-state energy and wavefunction, but you don't need to know anything about the excited states.  How should you go about choosing $mn$ pairs for Jacobi rotations?  What if instead of the ground state you want to find only a particular excited state?

\item For the harmonic oscillator example described above, compare the energy eigenvalues to the exact values that you would obtain from an analytic treatment of the quantum harmonic oscillator.  What is the main reason for the inaccuracies?  What change should you make to obtain greater accuracy?

\item Suppose that the particle in question is an electron and that $a = 1$~nm.  What, then, is the size of one natural unit of energy, in electron-volts?  (Hint:  What combination of $\hbar$, $M$, and $a$ has units of energy?)

\item Modify the code in Fig.~\ref{fig:MathematicaCode} to solve the ``quantum bouncer'' potential:  $V(x)=\infty$ for $x<0$ and $V(x) = \alpha x$ for $x>0$, where $\alpha$ is a constant for which a reasonable value, in natural units, is 500.  Explain why the initial Hamiltonian matrix for this system is qualitatively very different from that for the harmonic oscillator.  Note the final energy eigenvalues when \texttt{nMax} = 8, then check them by increasing \texttt{nMax} and either repeating the Jacobi rotation process or simply using the \texttt{Eigensystem} function.  Briefly discuss the resulting energy level structure and the shapes of the energy eigenfunctions.

\item Modify the code in Fig.~\ref{fig:MathematicaCode} to solve a symmetric double-well potential.  A simple way to do this is to use Mathematica's \texttt{If} function to set $V(x)$ to a positive constant within a narrow interval around $x=0.5$, and set $V(x)=0$ everywhere else.  Alternatively, you could devise a smooth double-well potential using a quartic polynomial.  Either way, make the central barrier high enough that several low-lying energy eigenvalues are below the barrier height.  Explain the general features of the initial Hamiltonian matrix, then carry out the solution and discuss the resulting eigenvalues and eigenfunctions.

\item Sometimes a Jacobi rotation will put the basis states into the ``wrong'' order, with a higher-energy state having a lower $n$ value than a lower-energy state.  (See Fig.~\ref{fig:MathematicaOutput} for an example.)  Write a Mathematica function, similar in many ways to \texttt{rotateCell}, that swaps the order of any two basis states, so that you can use this function at any time to put the states back into the ``right'' order.

\item Write a Mathematica function to automatically carry out an arbitrary number of successive Jacobi rotations, either zeroing out the largest remaining off-diagonal Hamiltonian matrix element during each step, or simply cycling through all the off-diagonal elements in order by row and column.  (It's a nice touch to use the \texttt{Pause} function to slow down the process, so you can watch the display update with each step.)  After testing your code, use it to explore the rate at which the Jacobi rotation algorithm converges.  That is, how many rotations per off-diagonal element are needed before all of these elements are smaller than a given threshold?

\end{enumerate}


\begin{thebibliography}{99}
\bibitem{remark1} Brute-force numerical integration of Eq.~(\ref{HamIP}) is feasible on personal computers for up to a few dozen rows and columns.  When larger matrices are needed, one should use a product-to-sum trigonometric identity to reduce the number of distinct integrals that must be evaluated.  Sometimes it is even possible to evaluate the integrals analytically.  See Ref.~\onlinecite{Marsiglio}.
\end{thebibliography}
\end{document}